\documentclass[12pt]{article}

\usepackage{scicite}

\usepackage{times}

\usepackage{amsmath}
\usepackage{amsfonts}
\usepackage{amssymb}
\usepackage{longtable}
\usepackage{graphicx}

\def\PKS{\hbox{PKS\,1830$-$211}}

\newenvironment{sciabstract}{%
\begin{quote} \bf}
{\end{quote}}

\newcounter{lastnote}

\title{\vspace{-30mm} A strong magnetic field in the jet base of a supermassive black hole}

\author
{Ivan Mart\'i-Vidal,$^{1\ast}$ S\'ebastien Muller,$^{1}$ Wouter Vlemmings,$^{1}$ \\
Cathy Horellou,$^{1}$ Susanne Aalto$^{1}$\\
\normalsize{$^{1}$ Department of Earth and Space Sciences, Chalmers University of Technology,} \\
\normalsize{Onsala Space Observatory, SE-43992, Onsala, Sweden} \\
\\
\normalsize{$^{\ast}$ To whom correspondence should be addressed; E-mail: mivan@chalmers.se} \vspace{10mm}
\\
{\em Published in Science, Volume 348, Issue 6232, pp. 311-314 (2015)}
}

\date{}

\begin{document} 


\baselineskip24pt


\maketitle


\begin{sciabstract}

Active galactic nuclei (AGN) host some of the most energetic phenomena in the Universe.
AGN are thought to be powered by accretion of matter onto a rotating disk that surrounds a supermassive black hole. Jet streams can be boosted in energy near the event horizon of the black hole and then flow outward along the rotation axis of the disk. The mechanism that forms such a jet and guides it over scales from a few light-days up to millions of light-years remains uncertain, but magnetic fields are thought to play a critical role. Using the Atacama large mm/submm array (ALMA), we have detected a polarization signal (Faraday rotation) related to the strong magnetic field at the jet base of a distant AGN, \PKS.
The amount of Faraday rotation (rotation measure) is proportional to the magnetic field strength along the line of sight times the density of electrons. Although it is impossible to precisely infer the magnetic fields in the region of Faraday rotation, the high rotation measures derived suggest magnetic fields of at least tens of Gauss (and possibly considerably higher) on scales of the order of light days (0.01 pc) from the black hole.

\end{sciabstract}

The AGN jets, related to the accretion mechanism in supermassive black holes, consist of relativistic plasma driven by strong and ordered magnetic fields. As a result of the magnetic interaction of the plasma, non-thermal (synchrotron) emission is produced.  
Studying the polarization of this non-thermal emission from AGN is a direct way to probe the structure and strength of magnetic fields in the vicinity of a black hole. Of particular importance is the observation of the rotation measure, $RM$, defined as the change of polarization angle as a function of wavelength squared. This quantity is directly related to the plasma density and the strength of the magnetic field along the line of sight.

To date, it has been extremely difficult to obtain accurate polarimetric information 
from the innermost (sub-parsec) regions of AGN; only emission at sub-millimeter wavelengths can escape from these regions, due to a large synchrotron self-absorption (SSA) that blocks the emission at longer wavelengths. Unfortunately, the sensitivity of polarization observations at sub-millimeter wavelengths has been so far strongly limited by the instrumentation.

Previous attempts to detect Faraday rotation at submillimeter (submm) wavelengths from AGN have yielded only upper limits \cite{Agudo,KuoM87} and marginal detections \cite{Trippe} that require strong assumptions about the absence of variability on timescales of weeks. There is a more robust detection for the Galactic center\cite{Marrone}, although the activity in this source is much lower than in AGN. Recently, measurements of Faraday rotation in the nearby AGN 3C\,84 (redshift $z=0.018$)  
have been reported at mm wavelengths \cite{plamb14}.

We have obtained measurements of Faraday rotation at frequencies up to 300\,GHz (about 1\,THz in the rest frame of the source) from \PKS, a powerful gravitationally-lensed AGN located at a redshift of $z = 2.5$ \cite{lid99}. At these frequencies, SSA is negligible in the whole jet of \PKS\ \cite{IMVPKS} and the maximum emission originates at the closest jet region to the black hole; the zone where the plasma is being injected and accelerated into the main jet stream. At lower frequencies, SSA hides this jet acceleration zone from view.  
These results are thus fundamental to better understand the role of magnetic fields in the AGN accretion and jet production, which are intimately related to the growth and evolution of supermassive black holes.

This detection has been possible thanks to the high resolution (sub-arcsecond) of our observations with ALMA, and to the use of a new differential polarimetry technique, which we present in \cite{SupportA} and briefly describe in the following lines.

The ALMA receivers detect the signal in two orthogonal linear polarizations, X and Y, where X is received from a horizontal dipole and Y from a vertical dipole in the frame of the antenna mount. 
The two lensed images of \PKS, which we call northeast (NE, upper-left in projection on sky) and southwest (SW, lower-right), are separated by $\sim$1''. In Fig.~1 we show an example of snapshot images in XX and YY of the two components of the gravitational lens, as well as their difference. The difference image contains information about the difference between NE and SW in Stokes parameters $Q$ and $U$. Our analysis makes use of the polarization ratio, $R_{pol}$, which is defined as

$$R_{pol} = \frac{1}{2}\left(\frac{R^{1,2}_{XX}}{R^{1,2}_{YY}}-1\right),$$

\noindent where $R_{XX}^{1,2}$ and $R_{YY}^{1,2}$ are the flux-density ratios between the two lensed images of the AGN, obtained separately from the XX and YY polarization products. $R_{pol}$ is a function of the parallactic angle of the antennas, $\psi$, and the observing wavelength, $\lambda$, and encodes information about the difference of polarization between the two images, via the approximately constant parameters $p_{dif}$ and $\alpha$ defined in \cite{SupportA}, as well as their rotation measure $RM$,

\begin{equation}
R_{pol} = p_{dif}\cos{\left( 2\phi'^0_1 -\alpha + 2\,RM(\lambda^2-\psi/RM) \right)},
\label{Eq2Fit}
\end{equation}

\noindent where $\phi'^0_1$ is the position angle of the polarization of image 1 at zero wavelength in the plane of the sky. The technique of differential polarimetry essentially enables estimation of $RM$ via fitting the observed sinusoidal dependence of $R_{pol}$ as a function of $\lambda^2$ and $\psi$, using Eq.~\ref{Eq2Fit}.

Our results are based on ALMA observations at sky frequencies 
around 100, 250, and 300\,GHz. Correcting for the cosmological redshift, these frequencies correspond to 350, 875, and 1050\,GHz in the frame of the source. More details on these observations, and a summary of the main goals of this ALMA project, can be found elsewhere \cite{mul14}. We also summarize all the observations in \cite{SupportC}. Our observations can be divided in two data-sets, one consisting of 6 epochs in 2012 (9 April to 16 June) and another of 9 epochs in 2014 (3 May to 27 August). 
In Fig.~2, we show the measured $R_{pol}$ between the two lensed images of \PKS. These measurements have been obtained from the $R_{XX}$ and $R_{YY}$ values fitted with the visibility-modelling software presented in \cite{UVMULTIFIT}. The uncertainties have been obtained with the standard error propagation approach, using the uncertainties in $R_{XX}$ and $R_{YY}$ that were derived from the covariance matrix of the visibility fitting, as described in \cite{UVMULTIFIT}.

The derivatives of $R_{pol}$ vs. $\lambda^2$, which are related to the $RM$ \cite{SupportA3},
are clearly different for different wavelength ranges. Between $\lambda^2 = 8$ and 12 mm$^2$, the maximum derivative is $4.4\times 10^{-3}$\,mm$^{-2}$, whereas between 0.8 and 1.6 mm$^2$ it is $70\times 10^{-3}$\,mm$^{-2}$. Because the maximum observed $R_{pol}$ are, in absolute value, similar at all wavelengths, the different derivatives of $R_{pol}$ vs. $\lambda^2$ must be due to larger $RM$ at shorter wavelengths (see \cite{SupportA3} for a more detailed discussion). Large variations of $RM$ with wavelength have been reported in other AGN \cite{Gabuzda5}, although at much longer wavelengths (cm), related to larger spatial scales in the jets.
Our finding cannot be explained easily if the $RM$ is only caused by an external (e.g. spherically symmetric) screen of material being accreted onto the black hole (as in the case of the $RM$ detected in the Galactic center \cite{Marrone}) and/or by external clouds. The size of the submm emitting region (estimated as the distance to the black hole at which the submm intensity is maximum) is only of the order of 0.01\,pc \cite{IMVPKS}. Hence, if the Faraday screen were extended and located far from the jet base, the rotation measure at submm wavelengths should not depend on the observing frequency, since the extent of the Faraday screen would be similar for all the submm jet emission. The Faraday screen must thus be close to the jet base, and change substantially on sub-parsec scales (Fig.~4).  
An increase of the $RM$ at shorter wavelengths would then be explained naturally as an increase of the magnetic field strength and/or electron density as we approach the black hole. Indeed, there are observations of other AGN at long wavelengths (cm) that show changes of $RM$ across the jets, both longitudinal and transversal\cite{Gabuzda1,Gabuzda2,Gabuzda3} that have been attributed to changes in particle density and magnetic fields in the jets, independent of a more distant external medium.

\begin{table}
\caption{Best-fit polarization values for the three epochs with quasi-simultaneous observations at 250 and 300\,GHz. $RM_{obs}$ are the rotation measures in the observer's frame and $RM_{true}$ are the rotation measures in the rest frame of the source. $RM_{true}$ is $(1+z)^2$ times larger than $RM_{obs}$.} 
\label{RMFitTab}
\begin{center}
\begin{tabular}{|c|c|c|c|}
\hline 
 & \multicolumn{3}{|c|}{{\bf Epoch } } \\ 
                         & 10 Apr 2012 & 23 May 2012 & 5 May 2014 \\ 
\hline 
$RM_{obs}$ (10$^6$ rad/m$^2$)  & 9.0$\pm$0.3 & 9.4$\pm$0.4 & 25.3$\pm$0.8 \\ 
\hline
$RM_{true}$ (10$^7$ rad/m$^2$)  & 11$\pm$0.4 & 11.5$\pm$0.5 & 31.2$\pm$1.0 \\ 
\hline 
$p_{dif}$ (10$^{-3}$)   & 12.6$\pm$0.4 & 3.8$\pm$0.3 & 3.5$\pm$0.3  \\ 
\hline 
$2\phi_0 - \alpha$ (deg) &   59$\pm$27  &  40$\pm$23  &  25$\pm$20   \\ 
\hline 
\end{tabular} 
\end{center}
\end{table}

We have three sets of observing epochs at 250 and 300\,GHz separated by a short time interval (1-2 days). In these three cases, we can directly estimate $RM$ and $p_{dif}$ by fitting $R_{pol}$ to the model given by Eq.~\ref{Eq2Fit}. The parameter estimates in these three datasets have been performed by least-squares minimization, comparing the measured $R_{pol}$ to the model predictions. The data at our lowest frequency band (i.e., 100\,GHz) have been discarded from the fit, since they trace different rotation measures from different regions of the jet, as we have already discussed. We show the fitting results in Fig.~3 and the estimated parameters in Table~1.  
Our estimated source-rest-frame $RM$s  are about two orders of magnitude higher than the highest values reported previously for other AGN, which are $\sim 10^6$~rad/m$^2$ \cite{Trippe,plamb14}.  

Although the two $RM$ measurements in 2012 are compatible, the estimate in 2014 is higher by more than a factor of two. Regarding the amplitude of $R_{pol}$, which is related to the fractional polarization and to the relative polarization angles among the NE and SW images, we find different values for the two observations in 2012. These two observations were serendipitously taken before and after a strong $\gamma$-ray flare, which had a very weak radio counterpart \cite{IMVPKS}. This leads us to speculate that the change in polarization may be correlated to the radio counterpart of that flare. Another $\gamma$-ray flare was detected in 2014 \cite{Ciprini2014}, also coincident with the time range of our 2014 observations. The new flare had a strong radio counterpart,  
which may also be related to the higher $RM$ that we measure in 2014. The high variability in $RM$ and $p_{dif}$, in connection to the $\gamma$-ray flaring events, points toward a roughly co-spatial origin of the $\gamma$-ray emission and the 250$-$300\,GHz rotation measures, hence favouring our interpretation of the $RM$ being caused at the region very close to the jet base.

The $RM$ is related to the line-of-sight integral of the electron density times the magnetic field, corrected for the cosmological redshift\cite{Bernet}. In units of rad/m$^2$,

\begin{equation}
RM = 8.1\times10^{5}\frac{1}{(1+z)^2}\int{n\,B_{||}dl},
\label{FarRotTotal}
\end{equation}

\noindent where $dl$ is the differential path along the line of sight (in pc), $z$ is the redshift ($z = 2.5$), $n$ is the particle density (in cm$^{-3}$) and $B_{||}$ the magnetic field projected in the line of sight (in Gauss). 
Our $RM$ will be an important test for detailed magnetohydrodynamical (MHD) models at the jet base, but such an analysis is beyond the scope of this paper. 
At a more basic level, it is impossible to unambiguously disentangle the contributions of the magnetic field, electron density and path length to the integral determining the rotation measure. This difficulty is exacerbated by the absence of direct information about the electron density or path length from observations, leading to the need to extrapolate from larger scales, which introduces additional uncertainty. The rotation measures derived here, $RM\simeq 10^8$~rad/m$^2$ in the rest frame of the source, are about a factor of $10^5$ greater than the rest-frame $RM$ 
values measured for parsec-scale AGN cores, where the derived magnetic fields have been independently measured to be $\simeq 0.05-0.10$~Gauss \cite{Gabuzda4,Gabuzda5}; this suggests that the magnetic fields in the sub-parsec regions we are probing are at least a few tens of Gauss, and possibly much higher. More exact estimates of these magnetic fields will require a separate dedicated study.

This is a clear indication of very high magnetic fields at the jet base, which should be dynamically important near the black hole and should in turn affect the accretion process. A similar conclusion was drawn from a statistical analysis of jet core-shifts from a complete sample of AGN, using high-resolution radio observations at centimeter wavelengths\cite{zaman14}. 
In the near future, our differential polarimetry technique can be used to further measure and monitor $RM$s at very short wavelengths, from this and other AGN. The monitoring of magnetic fields and particle densities at the closest jet regions to the black holes, via submm polarimetry, will allow us to study the tight connection between black-hole accretion and relativistic jets, the two fundamental pieces of the fascinating cosmic puzzle of Active Galactic Nuclei.

\clearpage

\begin{figure}[ht!]
\centering
\includegraphics[width=14cm]{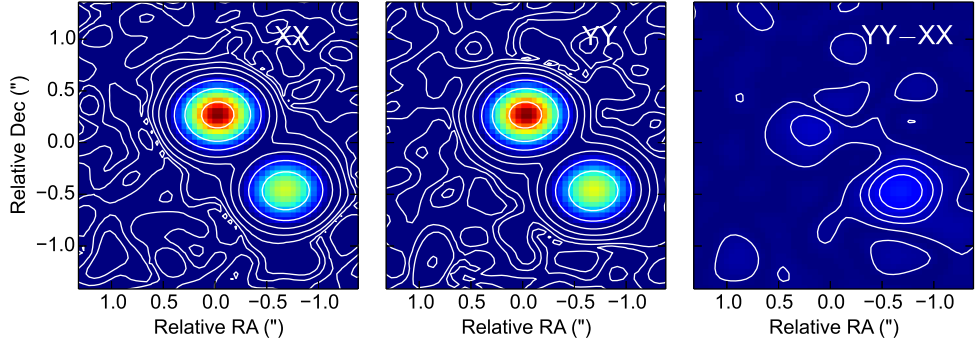}
\caption{ALMA image of the gravitationally-lensed blazar \PKS\ at 250\,GHz (875\,GHz in the source frame), taken on 30 June 2014. Left, in XX polarization; center, in YY polarization (with the peak normalized to that of the XX image); right is the difference between polarizations. Notice the small residual in the south-west lensed image, which encodes differential polarization information among the north-east and south-west images. The contours are set at 0.025, 0.05, 0.1, 0.2, 0.5, and 0.99 times the peak intensity.}
\label{Figure1}
\end{figure}

\begin{figure}[ht!]
\centering
\includegraphics[width=14cm]{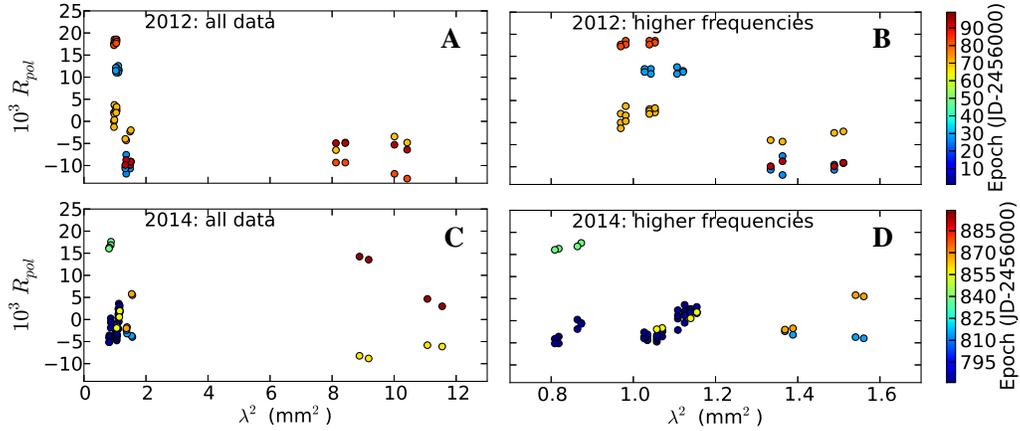}
\caption{Polarization ratio, $R_{pol}$, as a function of the wavelength squared, in the observer's frame, for all our ALMA observations. Left panels (A and C) are all data; right panels (B and D) are a zoom to the region of shorter wavelengths. The uncertainties, estimated from the post-fit covariance matrix as described in \cite{UVMULTIFIT}, are of the order of the symbol sizes.}
\label{Figure2}
\end{figure}

\begin{figure}[ht!]
\centering
\includegraphics[width=8cm]{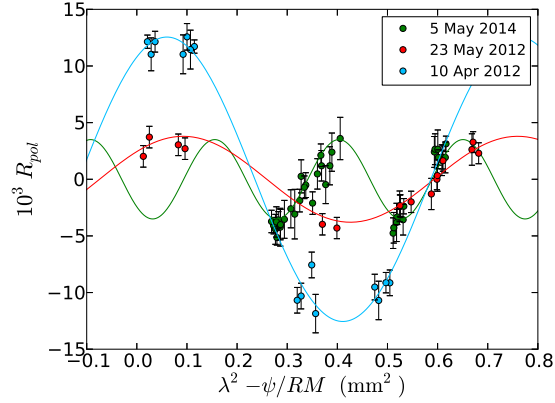}
\caption{Fits of our three epochs with quasi-simultaneous observations at 875 and 1050\,GHz (source rest frame) to the model given in Eq.~\ref{Eq2Fit}. We show $R_{pol}$ vs. $\lambda^2$ corrected by $-\psi/RM$, to obtain a sinusoidal behaviour.}
\label{Figure3}
\end{figure}

\begin{figure}[ht!]
\centering
\includegraphics[width=8cm]{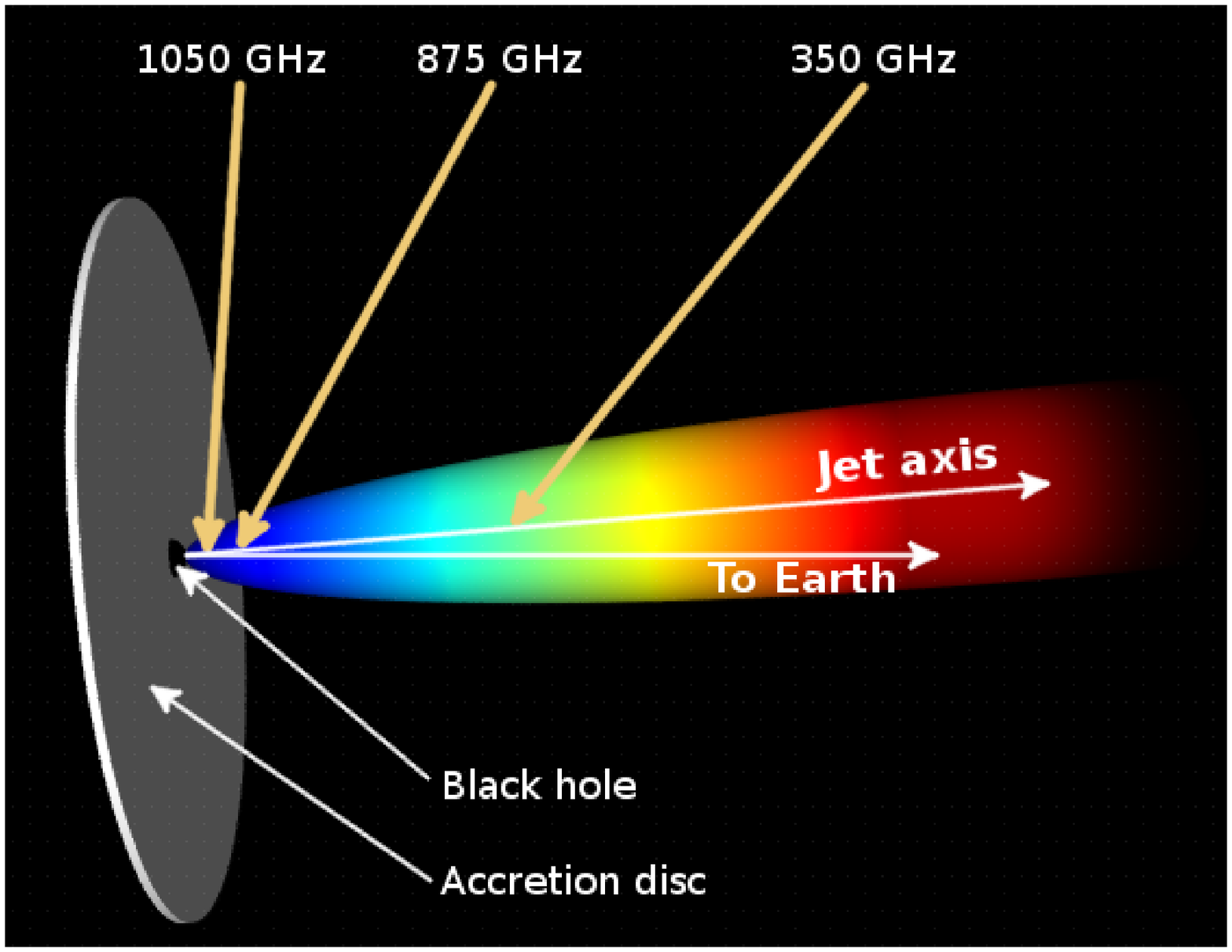}
\caption{Sketch of the jet launch/acceleration region in \PKS\ (not to scale). Emission at higher frequencies comes from material closer to the black hole, at sub-parsec scales. At these frequencies, the main contribution to the Faraday rotation, $RM$, must come from a zone close to the jet, in order to explain the different $RM$ between 350\,GHz and 0.8$-$1\,THz.}
\label{Figure4}
\end{figure}

\clearpage
\newpage




\clearpage


This paper makes use of the following ALMA data:
ADS/JAO.ALMA\#2011.0.00405.S, 
ADS/JAO.ALMA\#2012.1.00056.S, and 
ADS/JAO.ALMA\#2013.1.00020.S. 
ALMA is a partnership of ESO (representing
its member states), NSF (USA) and NINS (Japan), together with NRC
(Canada) and NSC and ASIAA (Taiwan), in cooperation with the Republic of
Chile. The Joint ALMA Observatory is operated by ESO, AUI/NRAO and NAOJ.


\clearpage

\clearpage

\renewcommand\thefigure{S\arabic{figure}}    
\setcounter{figure}{0}

\renewcommand\thetable{S\arabic{table}}    
\setcounter{table}{0}

\pagenumbering{arabic}

\begin{center}
\vspace{2mm}
\Huge{Supplementary materials for} \\
\vspace{3mm}
\Large{A strong magnetic field in the jet base of \\a supermassive black hole}\\ 
\vspace{5mm}
\normalsize{Ivan Mart\'i-Vidal, S\'ebastien Muller, Wouter Vlemmings, Cathy Horellou, Susanne Aalto \\
\vspace{2mm}
{\em Dpt. of Earth and Space Sciences, Chalmers Univ. of Technology,\\ Onsala Space Observatory (Sweden)}}

\end{center}

\section{Differential polarimetry}
\label{AppDifPol}

As we explain in ({\em 8}), the use of flux-density ratios among different components of an image allows us to overcome the less precise absolute flux calibration in the data analysis. Although the precision of the absolute flux calibration is typically worse than 5--10\%, an analysis based on the relative flux-densities of different parts of the same image allows us to detect signals with precisions several orders of magnitude higher.

We can apply the same strategy to the {\em polarimetry} signal (i.e., the brightness distribution of Stokes parameters), hence focusing our analysis on the {\em relative} polarization among different parts of the same image. In this section, we elaborate on the technique of {\em differential polarimetry} in interferometric observations, which allows us to exploit this possibility. 

Let us observe a source with a structure that can be split into a discrete set of point sources.
Let $k_i$ be the $k$-th Stokes parameter (i.e., $k$ can be either $I$, $Q$, $U$, or $V$) of the $i$-th point source. At a given observing epoch, and assuming that the point sources do not vary during the observation, the ratios of Stokes parameters between the sources have to remain constant, i.e. 

\begin{equation}
\frac{I_{i}}{I_{j}} = R^{ij}_I~;~\frac{Q_{i}}{Q_{j}} = R^{ij}_Q~;~\frac{U_{i}}{U_{j}} = R^{ij}_U~;~\frac{V_{i}}{V_{j}} = R^{ij}_V.
\end{equation}

Any change in the polarization of one source with respect to the other should map into changes in the $R^{ij}_k$ ratios. If the linear polarization  
of one of the sources changes, then $R^{ij}_Q$ and $R^{ij}_U$ will change accordingly. From now on, we assume that the $Q$ and $U$ parameters are given in the axis frame of the antenna mounts, and not in the sky frame. Hence, $Q$ and $U$ depend on the parallactic angle, $\psi$, of each observation.

Other factors that may affect, in principle, the $R^{ij}_Q$ and $R^{ij}_U$ ratios are polarization leakage in the antenna receivers and/or direction-dependent effects (DDEs) related, for instance, to beam-squint. The DDE effects should be very small, as long as the separation between the sources is small compared to the primary beam of the antennas. In our case, we consider two point sources corresponding to the two components of the
gravitational lens \PKS. The separation between these components (1 arcsec) is very small compared to the size of the antenna primary beam (which is 21 arcsec at 300\,GHz), and it is known from the last ALMA Call for Proposals that the effect of DDEs in the inner 1/3 of the ALMA antenna beams should be negligible.

\subsection{Differential polarimetry from dual-polarization observations}

We will now elaborate the polarimetric equations for the case of antennas with receiver feeds given in a linear polarization basis. The orthogonal axes are named X and Y, where X is received from a dipole in horizontal direction, in the frame of the antenna mount, and Y is received from a vertical dipole. 
Let us assume that our observations are performed in a mode in which the two orthogonal polarizations, X and Y, are recorded at each antenna, but only the XX and YY baseline products (i.e., not the XY and YX) are recovered. This mode is usually called {\em dual-polarization}, in contrast to the {\em full-polarization} mode where all the combinations, XX, YY, XY, and YX are computed.


In principle, it should not be possible to extract robust polarimetric information from the XX and YY products alone, since there may be small-amplitude calibration offsets between the X and Y signals in the antenna receivers that can introduce effects very similar to those from source polarization, especially in cases of short observations where the coverage of parallactic angle is small. However, we have the possibility of performing {\em differential} polarimetry among the source components in an image. This approach is inspired by the {\em Earth rotation polarimetry synthesis} technique (see, e.g., {\em 4}), although it has some essential differences. In differential polarimetry, we basically sacrifice some information related to the absolute polarization state of a source by means of extracting intra-field observables, which are, by construction, free of many calibration-related artifacts that limit the analysis based on ordinary polarimetric approaches.   

Let us call $R^{ij}_{XX}$ and $R^{ij}_{YY}$ the flux-density ratios between source components $i$ and $j$, obtained separately from the XX and YY visibility products, respectively. It is obvious that the {\em polarization ratio}, 

\begin{equation}
R_{pol} = \frac{1}{2}\left(\frac{R^{ij}_{XX}}{R^{ij}_{YY}}-1\right),
\label{RpolDef}
\end{equation}

\noindent is {\em independent} of any amplitude offset between the XX and YY products, so that it does encode robust polarimetric information inherent to the source. 

But what is the physical interpretation of the polarization ratio $R_{pol}$?
On the one hand, the XX products are related to the source brightness distribution of the sum of two Stokes parameters ({\em 19}) , $I+Q$, whereas the YY products are related to the brightness distribution of $I-Q$, i.e

\begin{equation}
R^{ij}_{XX} = \frac{(I+Q)_{i}}{(I+Q)_{j}} ~;~ R^{ij}_{YY} = \frac{(I-Q)_{i}}{(I-Q)_{j}}.
\label{PolRatEq1}
\end{equation} 

The polarization ratio will thus be

\begin{equation}
2R_{pol} + 1= \frac{R^{ij}_{XX}}{R^{ij}_{YY}} = \frac{(I+Q)_{i}}{(I-Q)_{i}}\times\frac{(I-Q)_{j}}{(I+Q)_{j}}.
\label{PolRatEq2}
\end{equation} 

Since typically $Q<<I$, we can approximate

\begin{equation}
2R_{pol}+1 = \frac{R^{ij}_{XX}}{R^{ij}_{YY}} \sim \left(1 + 2\left[\frac{Q}{I}\right]_{i} \right) \left(1 - 2\left[\frac{Q}{I}\right]_{j}\right) 
\label{PolRatEq3}
\end{equation}

The fractional linear polarization is defined as $p = \sqrt{Q^2+U^2}/I$, but it is not possible to extract information about the $U$ Stokes component from our dual-polarization visibility products. We can, however, constrain a minimum value for $p$, we call it $p^{Q} = Q/I$, so that the fractional polarization must be higher (or equal) to $p^{Q}$ in absolute value. The actual difference between $p$ and $p^{Q}$ in each source component will depend on its polarization angle, and how it aligns to the antenna axis. Hence, a change in $p^{Q}$ can be explained as either a change in the fractional polarization, $p$, or a change in the polarization angle, $\phi = \frac{1}{2}\tan^{-1}{(U/Q)}$, which maps into a different projection of the polarization vector in the direction where $Q$ is defined. With this definition of $p^{Q}$, we can rewrite Eq. \ref{PolRatEq3} as

\begin{equation}
2R_{pol}+1 \sim \left(1 + 2p^{Q}_{i} \right) \left(1 - 2p^{Q}_{j}\right) \sim 1 + 2\left(p^{Q}_{i}-p^{Q}_{j}\right),
\label{PolRatEq4}
\end{equation} 

\noindent which relates the minimum fractional linear polarization in the $i$ source to that of the $j$ source. It is straightforward to show that $p^{Q} = p\,\cos{(2\phi)}$, where $\phi$ is the polarization angle with respect to the X axis of the antennas. Hence,

\begin{equation}
R_{pol} = p_{i}\cos{\left(2\phi_{i}\right)} - p_{j}\cos{\left(2\phi_{j}\right)},
\label{PolRatEq5}
\end{equation}

\noindent where $p_{i}$ and $p_{j}$ are the fractional polarizations of the $i$-th and $j$-th sources, respectively, and $\phi_i$ and $\phi_j$ are their polarization angles. 

From Eq. \ref{PolRatEq5}, it is easy to conclude that the $R_{pol}$ can only be zero (i.e., no signal of differential polarization) either if both $p_{i}$ and $p_{j}$ are zero (i.e., there is no polarization in any source component) {\em or} if the polarization angles are such that the ratio of their double sines equals the ratio of the fractional polarizations. However, the latter situation is unlikely to happen (the chance for the antennas to be just on the right parallactic angle that cancels the right-hand side of Eq. \ref{PolRatEq5} is small) and it can be ruled out in the cases when $R_{pol}$ is measured at {\em different} parallactic angles.

\subsection{Effect of Faraday rotation}
\label{DiffFaraday}

Robust information on the strength of the magnetic field and the electron density, either from the jet or from the accretion flow, can be extracted from the dependence of polarization angle with observing wavelength; the so called {\em Faraday rotation},

\begin{equation}
\phi = \phi_0 + RM\,\lambda^2,
\label{FarRot}
\end{equation}

\noindent where $\phi$ is the polarization angle, $\phi_0$ is the angle at zero wavelength, $RM$ is the rotation measure, and $\lambda$ is the wavelength.  
If the polarization angle of each source changes with wavelength, due to Faraday rotation, then 

\begin{equation}
R_{pol} = p_{i}\cos{\left(2(\phi^0_{i} + RM_{i}\lambda^2)\right)} - p_{j}\cos{\left(2(\phi^0_{j} + RM_{j}\lambda^2)\right)}.
\label{FarRotEq1}
\end{equation}

We can rearrange terms in Eq. \ref{FarRotEq1} by defining

\begin{equation}
\Delta = 2\left(\phi^0_j-\phi^0_i+(RM_j-RM_i)\lambda^2\right)
\label{DefsEq1}
\end{equation}

\noindent and

\begin{equation}
p_{dif} = \sqrt{p_i^2+p_j^2-2p_ip_j\cos{\Delta}}~~;~~\alpha = \arctan{\left(\frac{p_j\sin{\Delta}}{p_i-p_j\cos{\Delta}}\right)},
\label{DefsEq2}
\end{equation} 

\noindent so that

\begin{equation}
R_{pol} = p_{dif}\cos{\left( 2\phi^0_i + 2\,RM_i\lambda^2-\alpha \right)}.
\label{FarRotEq2}
\end{equation} 

The angles $\phi^0_i$ and $\phi^0_j$ are given in the frame of the antenna axes. Hence, to compute them in the sky frame, 
we have to account for the axes rotation of the antennas, via the parallactic angle $\psi$. 
Basically, the data must be divided into segments of roughly constant $\psi$, so that its effect can be subtracted from $\phi^0_i$ in all the $R_{pol}$ estimates. Taking the parallactic angle into account, Eq. \ref{FarRotEq2} becomes

\begin{equation}
R_{pol} = p_{dif}\cos{\left( 2\phi'^0_i + 2\,RM_i\lambda^2-\alpha-2\psi \right)},
\label{FarRotEq3}
\end{equation} 

\noindent where now $\phi'^0_i$ is the position angle of the polarization of source $i$ at infinite frequency in the frame of the sky and $\psi$ is the parallactic angle at which $R_{pol}$ has been measured.

\subsection{Special cases}
\label{AppDifPolMax}

Equation \ref{FarRotEq3} may have a complicated dependence on $\lambda^2$, via the $\alpha$ and $\Delta$ parameters. However, there are special cases where the equation can be simplified notably. In this section, we focus on two of these special cases.

\begin{itemize}

\item {\bf Gravitational lens}

If the two sources, $i$ and $j$, are lensed images of the same background source (i.e., we are observing a gravitational lens), their rotation measures should be similar, as well as the wavelength dependence of their fractional polarizations.  Then, $\Delta$ and $\alpha$ will roughly be constants (see Eqs. \ref{DefsEq1} and \ref{DefsEq2}) and Eq. \ref{FarRotEq2} will simplify $R_{pol}$ as a pure cosine with an argument equal to $\lambda^2 - \psi/RM$, a frequency term equal to $2\,RM$, and a phase offset equal to $2\phi'^0_i - \alpha$, i.e

\begin{equation}
R_{pol} = p_{dif}\cos{\left( (2\phi'^0_i - \alpha) + 2\,RM_i(\lambda^2 - \psi/RM_i) \right)}.
\label{FarRotCorr}
\end{equation}

In a gravitational lens, the time delay between the images, together with the intrinsic source variability, will introduce a rotation between the polarization angles $\phi_i$ and $\phi_j$. In addition, if the source is extended and its polarization structure is not uniform, differential amplification through the source structure (which is different for each image) will also introduce an effective rotation between the polarization vectors of the images. As a result, either source variability and/or a non-uniform polarization structure will make both $\Delta$ and $p_{dif}$ (see Eqs.~\ref{DefsEq1} and \ref{DefsEq2}) different from zero.



If $R_{pol}$ is measured at random values of $\lambda$ and/or $\psi$, it is reasonable to assume that the maximum observed $R_{pol}$, in absolute value, should be close to the maximum theoretical value of $R_{pol}$ derived from Eq. \ref{FarRotEq3}, i.e. $(R_{pol})^{max} \sim p_{dif}$. 
This assumption does not account for the time variation of $p_{dif}$, which is as large as a factor of several (see Table~1), but should be approximate enough for an order-of-magnitude discussion.

Since $RM_i \sim RM_j \sim RM$, we can differentiate Eq. \ref{FarRotEq3} with respect to $\lambda^2$ easily,

\begin{equation}
R'_{pol}(\lambda) = \frac{d\,R_{pol}}{d\,\lambda^2} = -2\,p_{dif}RM\,\sin{\left( 2\phi'^0_i + 2\,RM\lambda^2-\alpha-2\psi \right)},
\label{FarRotEq4}
\end{equation} 

\noindent where we assume that $p_{dif}$ and $\alpha$ are nearly independent of $\lambda$. We now assume that the maximum observed derivatives of $R_{pol}$ vs. $\lambda^2$ should not be far to the maximum theoretical value of the derivative, $(R'_{pol})^{max} \sim 2\,p_{dif}RM$. Indeed, using the $p_{dif}$ values reported in Table \ref{RMFitTab}, and the maximum $R'_{pol}$ observed at the shortest wavelengths, we derive $RM$ from 6$\times$10$^6$ to 18$\times$10$^6$\,rad/m$^2$, which are not far from the estimated values given in Table \ref{RMFitTab}. The ratio of maximum derivatives 
observed at two different wavelengths will thus be

\begin{equation}
\frac{\left(R'_{pol}(\lambda_1)\right)^{max}}{\left(R'_{pol}(\lambda_2)\right)^{max}} \sim \frac{RM(\lambda_1)}{RM(\lambda_2)}.
\label{FarRotEq5}
\end{equation} 

This expression is based on rough estimates and should be taken with care. However, for large variations of $(R'_{pol})^{max}$ at different wavelengths, it allows us to conclude that the $RM$ must also vary, in a similar way, between the different wavelengths.

\item {\bf Unpolarized source}

If only one of the two sources, $j$, is polarized (so $i$ is not polarized), then Eq. \ref{PolRatEq5} reduces to $R_{pol} = p_j\,\cos{\left(2\,\phi_j\right)}$. Thus, the $R_{pol}$ ratio encodes information on the absolute polarization state of source $j$.

\end{itemize}

\subsection{Polarization leakage in the XX and YY products}

The estimate of $R_{pol}$ is, by construction, insensitive to bandpass and/or gain differences among the X and Y signals.
The only calibration effects that can change the value of $R_{pol}$ is polarization leakage in the antenna receivers (plus any cross-polarization delay or phase). However, the effect of polarization leakage on the XX and YY products is very small (second-order corrections), compared to its effect on the XY and YX products (first-order corrections). 

Following ({\em 19}), the Jones matrix for the correction of leakage and cross-phase (or cross-delay) among the X and Y signals in an antenna is

\begin{equation}
J = \begin{bmatrix}1 & 0 \\0 & K\end{bmatrix}\times\begin{bmatrix}1 & D_x \\D_y & 1\end{bmatrix} = \begin{bmatrix}1 & D_x \\D_yK & K\end{bmatrix},
\label{JonesAll}
\end{equation}

\noindent where $K$ is a phase-like factor and $D_x$ and $D_y$ are the complex D-terms that model the polarization leakage in the antenna receivers.  
Since we do not have cross-polarization products in dual-polarization observations, it is not possible to solve for $K$ in the calibration (i.e., to separate it from the phase gains), but, in any case, its effects on the dual-polarization visibilities will not be different from a phase added to the YY product. The observed XX and YY visibilities, for the baseline formed by a pair of antennas $A$ and $B$, will be:

\begin{equation}
V_{xx}^{obs} = (D^A_x V_{yy}+V_{xy})(D^*)^B_y + D^A_xV_{yx}+V_{xx}
\label{VxxObs}
\end{equation}

and 

\begin{equation}
V_{yy}^{obs} = \left((D^A_y V_{xx} + V_{yx})(D^*)^B_x + D^A_yV_{xy}+V_{xx}\right)K^A(K^*)^B.
\label{VyyObs}
\end{equation}

It is clearly seen that Eq. \ref{VxxObs} and \ref{VyyObs} are basically symmetric one to the other, with the exception of a global phase-like factor, $K^A(K^*)^B$ (i.e., the difference of X-Y cross-delays among antennas $A$ and $B$) that will be fully absorbed in the ordinary phase-gain calibration of the YY visibilities. Hence, only the antenna leakage (and not the X-Y delay) may introduce differences between the source images in XX and YY. 

\begin{figure}[ht!]
\centering
\includegraphics[width=13cm]{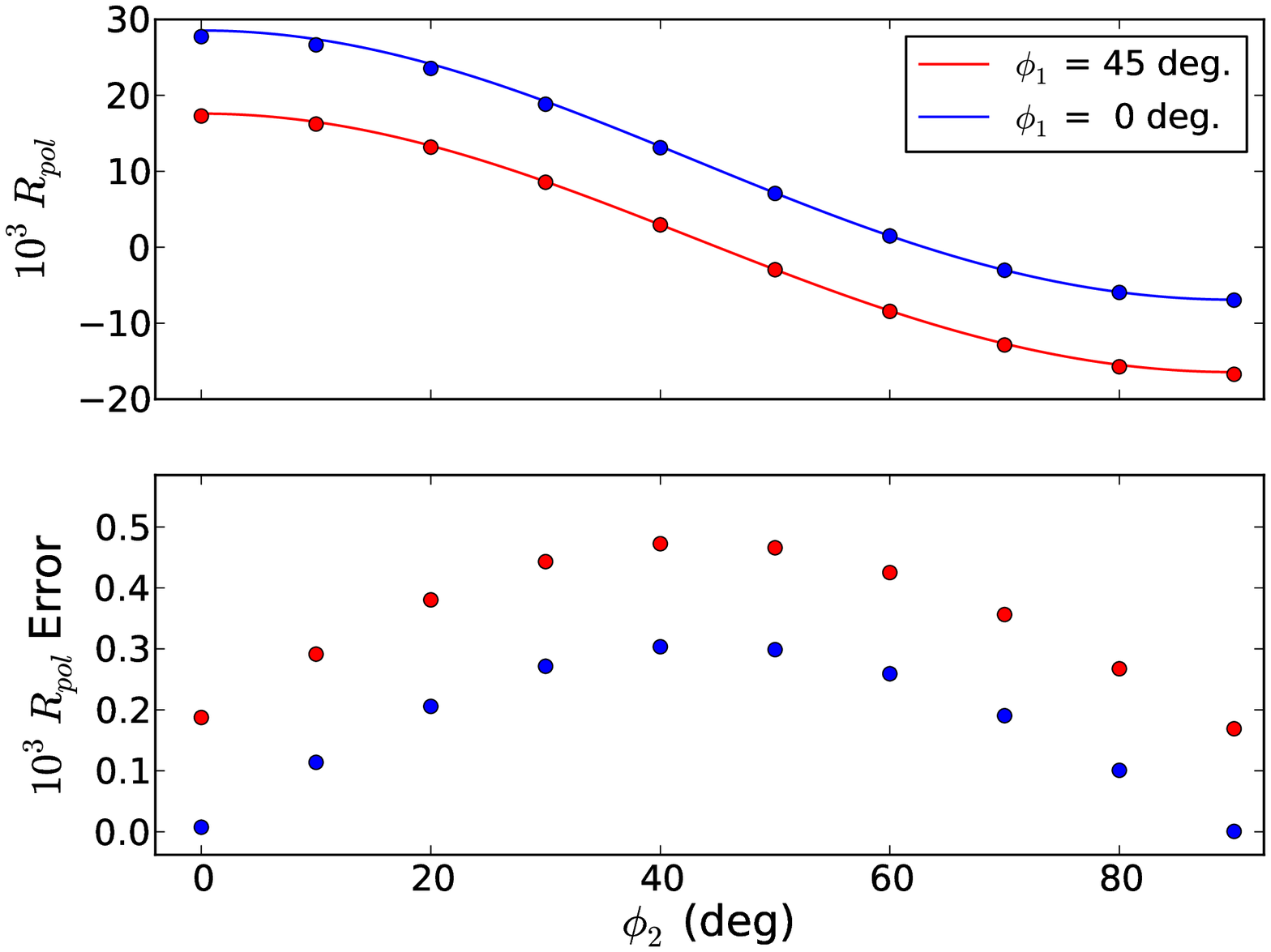}
\caption{Simulated $R_{pol}$, taking into account effects from polarization leakage in the antenna receivers. Top panel, fitted $R_{pol}$ values (circles) and the model predictions computed from Eq.~\ref{PolRatEq5} (solid lines). Bottom, difference in $R_{pol}$ between the fitted values and the model predictions.}
\label{Simulations}
\end{figure}

We have simulated interferometric observations of two polarized sources, located in the same field of view, and checked the robustness of $R_{pol}$ against polarization leakage in the antenna receivers. The two sources are separated by two synthesized beams, their intensity ratio is set to 1.4, and their linear polarizations are 1\% (with position angle, $\phi_1$, of either 0 or 45 degrees) and 1.7\% (with position angle, $\phi_2$, from 0 to 90 degrees, in steps of 10 degrees), respectively. No Faraday rotation has been introduced in the sources. We have simulated 10 antennas with leakages of similar amplitudes and random phases. The amplitude leakage used is 1\% (for ALMA, the leakage level in the antenna receivers is of the order of 1\%, at most). All simulated results are shown in Fig.~\ref{Simulations}, where it can be seen that all measured values of $R_{pol}$ follow the prediction of Eq.~\ref{FarRotEq3}. It is also clear that $R_{pol}$ is quite insensitive to the antenna leakage. All errors are below 5$\times$10$^{-4}$. The maximum errors are obtained when the $U$ Stokes signal is maximum in both sources (i.e., the polarization angles of both sources are close to 45 degrees). This is an expected result, since the leakage from $V_{xy}$ and $V_{yx}$ into $V_{xx}$ and $V_{yy}$ is larger for higher signals in U Stokes.

In these simulations, we have used a random phase distribution between 0 and 45 degrees for the D-terms, adding a correlation between $D_x$ and $D_y$ at each antenna (ideally, the phases of $D_x$ should be shifted by roughly 180 degrees with respect to those of $D_y$). Even if there was no correlation between $D_x$ and $D_y$, lower errors in $R_{pol}$ would be obtained. 
Unfortunately, there are not enough full-polarization ALMA observations reported yet, so it is difficult to constrain the true phase dispersion of antenna D-terms for ALMA.

\section{Epochs of observation and measurements}
\label{Measurements}

In this section, we give a list of all our observing epochs, showing the dates and times of the observations and the center observing frequencies. The bandwidth in each frequency window is $\sim2$\,GHz. We also show the measured flux-density ratios, as derived using the visibility-fitting program described in ({\em 10}). This program allows us to fit arbitrary combinations of geometrical models to the source structure, by fitting the models directly to the interferometric observables (i.e., the visibilities), hence bypassing the image deconvolution process. All the artifacts that could potentially be generated by the image deconvolution and reconstruction are thus avoided. In ({\em 10}), the reader will find a deeper discussion about the advantages of visibility fitting (compared to imaging) with a special emphasis on source structures like that of \PKS.

\begin{center}
\begin{longtable}{cccccccc}
\caption[List of all our ALMA observations and results on 2012]{List of all our ALMA observations and results on 2012. Column 2 is the time range of the observations, column 3 are the center frequencies, column 4 are the flux densities of the North-East component, column 5 are the flux-density ratios of Stokes I (i.e., adding the visibility products in XX and YY), columns 6 and 7 are the flux ratios in XX and YY, respectively, and column 8 are the ratios of differential polarization. Numbers in parenthesis at the right of each quantity are the uncertainties of the last significant figures.} 
\label{tab:contdata-c0} \\

\hline 
Date & Time  & Freq  & $f_{NE}$ & $R_{XX+YY}$ & $R_{XX}$ & $R_{YY}$ & $R_{pol}\times 10^3$ \\ 
     & (UTC)    & (GHz) & (Jy)     &       &            &            &             \\  
     \hline
\endfirsthead

\hline \multicolumn{8}{|r|}{{-- continued from previous page}} \\ \hline
Date & Time  & Freq  & $f_{NE}$ & $R_{XX+YY}$ & $R_{XX}$ & $R_{YY}$ & $R_{pol}\times 10^3$ \\ 
     & (UTC)    & (GHz) & (Jy)     &       &            &            &             \\  
     \hline
\endhead

\hline \multicolumn{8}{|r|}{{Continued on next page}} \\ \hline
\endfoot

\hline \hline
\endlastfoot

       09\,Apr\,2012&        06:23--06:57&245.9&0.7&1.238\,(2)&1.214\,(3)&1.262\,(3)& -18.905\,(59)\\
&&244.1&&1.237\,(2)&1.214\,(2)&1.258\,(2)& -17.328\,(44)\\
&&257.0&&1.232\,(2)&1.197\,(3)&1.265\,(3)& -26.730\,(81)\\
&&259.7&&1.232\,(2)&1.202\,(2)&1.261\,(2)& -23.441\,(64)\\
       09\,Apr\,2012&        07:42--08:16&245.9&0.8&1.237\,(2)&1.213\,(2)&1.261\,(2)& -19.068\,(40)\\
&&244.1&&1.234\,(1)&1.205\,(2)&1.260\,(1)& -21.818\,(38)\\
&&257.0&&1.236\,(2)&1.216\,(2)&1.256\,(2)& -15.883\,(33)\\
&&259.7&&1.239\,(1)&1.214\,(2)&1.264\,(2)& -19.429\,(37)\\
       11\,Apr\,2012&        06:06--06:59&283.3&0.6&1.224\,(1)&1.252\,(1)&1.195\,(1)&  23.933\,(41)\\
&&285.2&&1.224\,(1)&1.253\,(2)&1.193\,(2)&  25.270\,(47)\\
&&293.8&&1.222\,(1)&1.251\,(2)&1.194\,(1)&  23.744\,(42)\\
&&296.0&&1.223\,(1)&1.252\,(1)&1.194\,(1)&  24.158\,(40)\\
       11\,Apr\,2012&        07:47--08:40&283.3&0.7&1.250\,(3)&1.277\,(3)&1.223\,(3)&  22.080\,(76)\\
&&285.2&&1.252\,(3)&1.279\,(3)&1.224\,(3)&  22.633\,(78)\\
&&293.8&&1.251\,(2)&1.278\,(3)&1.224\,(2)&  22.137\,(60)\\
&&296.0&&1.251\,(2)&1.280\,(2)&1.223\,(2)&  23.430\,(62)\\
       22\,May\,2012&        09:23--10:00& 92.9&1.8&1.444\,(2)&1.431\,(1)&1.457\,(1)&  -8.650\,( 7)\\
&& 94.8&&1.441\,(2)&1.430\,(1)&1.453\,(1)&  -7.950\,( 7)\\
&&103.4&&1.442\,(2)&1.427\,(1)&1.457\,(1)& -10.362\,( 9)\\
&&105.3&&1.445\,(2)&1.429\,(1)&1.462\,(1)& -11.556\,(11)\\
       23\,May\,2012&        04:38--05:14&245.9&0.9&1.495\,(2)&1.487\,(2)&1.504\,(2)&  -5.619\,(11)\\
&&244.1&&1.495\,(1)&1.486\,(2)&1.503\,(2)&  -5.722\,( 9)\\
&&257.0&&1.499\,(2)&1.488\,(2)&1.510\,(2)&  -7.254\,(14)\\
&&259.7&&1.499\,(2)&1.488\,(2)&1.509\,(2)&  -7.090\,(12)\\
       23\,May\,2012&        05:47--06:21&302.8&0.8&1.489\,(2)&1.500\,(2)&1.478\,(2)&   7.306\,(16)\\
&&304.7&&1.489\,(2)&1.493\,(2)&1.483\,(2)&   3.371\,( 7)\\
&&292.4&&1.490\,(2)&1.499\,(2)&1.481\,(2)&   6.010\,(11)\\
&&294.3&&1.489\,(2)&1.499\,(2)&1.480\,(2)&   6.420\,(14)\\
       23\,May\,2012&        09:14--09:51&302.8&0.8&1.532\,(2)&1.537\,(2)&1.526\,(2)&   3.769\,( 9)\\
&&304.7&&1.534\,(2)&1.537\,(2)&1.531\,(2)&   1.894\,( 4)\\
&&292.4&&1.528\,(2)&1.533\,(2)&1.524\,(2)&   2.822\,( 6)\\
&&294.3&&1.531\,(2)&1.537\,(3)&1.523\,(3)&   4.562\,(11)\\
       23\,May\,2012&        10:27--11:04&302.8&0.8&1.539\,(2)&1.543\,(3)&1.535\,(3)&   2.410\,( 6)\\
&&304.7&&1.540\,(2)&1.541\,(3)&1.540\,(3)&   0.292\,( 1)\\
&&292.4&&1.531\,(2)&1.536\,(2)&1.526\,(2)&   3.146\,( 7)\\
&&294.3&&1.535\,(3)&1.539\,(3)&1.530\,(3)&   2.810\,( 7)\\
       04\,Jun\,2012&        07:18--07:52&302.8&0.7&1.259\,(1)&1.303\,(2)&1.214\,(1)&  36.314\,(62)\\
&&304.7&&1.260\,(1)&1.301\,(2)&1.216\,(1)&  35.068\,(61)\\
&&292.4&&1.258\,(1)&1.301\,(1)&1.214\,(1)&  35.756\,(57)\\
&&294.3&&1.258\,(1)&1.301\,(2)&1.214\,(1)&  35.750\,(58)\\
       04\,Jun\,2012&        08:32--09:07&302.8&0.7&1.262\,(2)&1.305\,(2)&1.218\,(2)&  35.767\,(64)\\
&&304.7&&1.262\,(2)&1.304\,(2)&1.219\,(2)&  34.876\,(63)\\
&&292.4&&1.260\,(1)&1.303\,(2)&1.217\,(1)&  35.289\,(60)\\
&&294.3&&1.261\,(1)&1.303\,(2)&1.218\,(1)&  34.970\,(60)\\
       04\,Jun\,2012&        09:42--10:18& 92.9&1.7&1.327\,(2)&1.304\,(1)&1.350\,(1)& -17.221\,(17)\\
&& 94.8&&1.319\,(2)&1.292\,(1)&1.347\,(1)& -20.342\,(20)\\
&&103.4&&1.308\,(2)&1.276\,(1)&1.340\,(1)& -23.733\,(24)\\
&&105.3&&1.306\,(2)&1.273\,(1)&1.340\,(1)& -24.927\,(27)\\
       15\,Jun\,2012&        07:18--07:54&245.9&0.8&1.292\,(1)&1.266\,(1)&1.317\,(1)& -19.511\,(26)\\
&&244.1&&1.290\,(1)&1.263\,(1)&1.315\,(1)& -19.585\,(26)\\
&&257.0&&1.291\,(1)&1.266\,(1)&1.314\,(1)& -18.076\,(28)\\
&&259.7&&1.292\,(1)&1.265\,(1)&1.317\,(1)& -19.520\,(28)\\
       15\,Jun\,2012&        08:53--09:29& 92.9&1.7&1.313\,(2)&1.303\,(1)&1.324\,(1)&  -8.193\,( 7)\\
&& 94.8&&1.308\,(2)&1.296\,(1)&1.321\,(1)&  -9.428\,( 8)\\
&&103.4&&1.304\,(2)&1.288\,(1)&1.320\,(1)& -12.010\,(10)\\
&&105.3&&1.301\,(2)&1.281\,(1)&1.320\,(1)& -14.620\,(13)\\
\hline 
\end{longtable} 
\end{center}

\begin{center}
\begin{longtable}{cccccccc}
\caption[List of all our ALMA observations and results on 2014]{List of all our ALMA observations and results on 2014. Column 2 is the time range of the observations, column 3 are the center frequencies, column 4 are the flux densities of the North-East component, column 5 are the flux-density ratios of Stokes I (i.e., adding the visibility products in XX and YY), columns 6 and 7 are the flux ratios in XX and YY, respectively, and column 8 are the ratios of differential polarization. Numbers in parenthesis at the right of each quantity are the uncertainties of the last significant figures.} 
\label{tab:contdata-c1} \\

\hline 
Date & Time  & Freq  & $f_{NE}$ & $R_{XX+YY}$ & $R_{XX}$ & $R_{YY}$ & $R_{pol}\times 10^3$ \\ 
     & (UTC)    & (GHz) & (Jy)     &       &            &            &             \\  
     \hline
\endfirsthead

\hline \multicolumn{8}{|r|}{{-- continued from previous page}} \\ \hline
Date & Time  & Freq  & $f_{NE}$ & $R_{XX+YY}$ & $R_{XX}$ & $R_{YY}$ & $R_{pol}\times 10^3$ \\ 
     & (UTC)    & (GHz) & (Jy)     &       &            &            &             \\ 
     \hline 
\endhead

\hline \multicolumn{8}{|r|}{{Continued on next page}} \\ \hline
\endfoot

\hline \hline
\endlastfoot

       03\,May\,2014&        09:46--10:16&321.0&0.6&1.931\,(2)&1.921\,(3)&1.940\,(3)&  -4.948\,(10)\\
&&322.8&&1.932\,(3)&1.926\,(4)&1.937\,(4)&  -2.813\,( 8)\\
&&331.6&&1.944\,(2)&1.937\,(2)&1.952\,(2)&  -3.689\,( 6)\\
&&333.5&&1.946\,(2)&1.939\,(2)&1.954\,(2)&  -3.915\,( 6)\\
       05\,May\,2014&        05:23--05:48&282.9&0.5&2.117\,(3)&2.113\,(3)&2.122\,(3)&  -2.215\,( 5)\\
&&285.0&&2.128\,(2)&2.117\,(3)&2.140\,(3)&  -5.305\,(11)\\
&&295.0&&2.141\,(3)&2.133\,(4)&2.150\,(4)&  -3.930\,(10)\\
&&296.1&&2.139\,(2)&2.132\,(3)&2.147\,(3)&  -3.354\,( 6)\\
       05\,May\,2014&        07:40--08:05&282.9&0.5&2.100\,(4)&2.090\,(5)&2.111\,(5)&  -4.998\,(16)\\
&&285.0&&2.098\,(3)&2.091\,(4)&2.107\,(4)&  -3.702\,(10)\\
&&295.0&&2.120\,(4)&2.113\,(5)&2.127\,(5)&  -3.127\,(11)\\
&&296.1&&2.117\,(3)&2.109\,(4)&2.125\,(4)&  -3.718\,( 9)\\
       05\,May\,2014&        07:40--08:05&282.9&0.5&2.091\,(4)&2.083\,(5)&2.100\,(5)&  -4.119\,(13)\\
&&285.0&&2.098\,(3)&2.097\,(4)&2.100\,(4)&  -0.572\,( 2)\\
&&295.0&&2.116\,(4)&2.112\,(5)&2.120\,(5)&  -2.004\,( 7)\\
&&296.1&&2.115\,(3)&2.107\,(4)&2.122\,(4)&  -3.604\,( 9)\\
       05\,May\,2014&        07:40--08:05&282.9&0.5&2.081\,(4)&2.079\,(5)&2.084\,(6)&  -1.151\,( 4)\\
&&285.0&&2.092\,(4)&2.096\,(5)&2.089\,(5)&   1.628\,( 6)\\
&&295.0&&2.110\,(5)&2.109\,(6)&2.112\,(7)&  -0.710\,( 3)\\
&&296.1&&2.110\,(4)&2.110\,(5)&2.110\,(5)&  -0.095\,( 0)\\
       05\,May\,2014&        08:40--09:05&279.2&0.5&2.069\,(3)&2.064\,(4)&2.075\,(4)&  -2.530\,( 7)\\
&&281.1&&2.072\,(3)&2.071\,(4)&2.072\,(4)&  -0.217\,( 1)\\
&&290.0&&2.090\,(3)&2.089\,(4)&2.092\,(4)&  -0.741\,( 2)\\
&&291.8&&2.097\,(4)&2.092\,(4)&2.102\,(4)&  -2.355\,( 7)\\
       05\,May\,2014&        08:40--09:05&279.2&0.5&2.073\,(3)&2.071\,(3)&2.076\,(3)&  -1.156\,( 3)\\
&&281.1&&2.074\,(3)&2.075\,(4)&2.073\,(4)&   0.265\,( 1)\\
&&290.0&&2.093\,(3)&2.092\,(3)&2.094\,(3)&  -0.310\,( 1)\\
&&291.8&&2.097\,(3)&2.093\,(4)&2.102\,(4)&  -2.069\,( 5)\\
       05\,May\,2014&        08:40--09:05&279.2&0.5&2.074\,(4)&2.071\,(4)&2.077\,(4)&  -1.372\,( 4)\\
&&281.1&&2.074\,(4)&2.074\,(5)&2.074\,(5)&   0.000\,( 0)\\
&&290.0&&2.096\,(4)&2.092\,(4)&2.100\,(4)&  -1.738\,( 5)\\
&&291.8&&2.099\,(4)&2.096\,(5)&2.101\,(5)&  -1.142\,( 4)\\
       05\,May\,2014&        10:15--10:39&279.2&0.6&2.078\,(2)&2.081\,(2)&2.077\,(2)&   0.818\,( 1)\\
&&281.1&&2.080\,(2)&2.080\,(2)&2.082\,(2)&  -0.336\,( 0)\\
&&290.0&&2.099\,(2)&2.100\,(2)&2.100\,(2)&  -0.071\,( 0)\\
&&291.8&&2.103\,(2)&2.103\,(2)&2.104\,(2)&  -0.261\,( 0)\\
       06\,May\,2014&        08:53--09:23&321.0&0.7&2.106\,(2)&2.093\,(3)&2.119\,(3)&  -6.181\,(11)\\
&&322.8&&2.108\,(3)&2.092\,(3)&2.124\,(3)&  -7.508\,(17)\\
&&331.6&&2.122\,(2)&2.106\,(2)&2.138\,(2)&  -7.298\,(11)\\
&&333.5&&2.125\,(2)&2.112\,(2)&2.138\,(2)&  -5.988\,( 8)\\
       06\,Jun\,2014&        07:18--07:35&240.2&1.2&1.872\,(2)&1.857\,(2)&1.888\,(2)&  -8.236\,(11)\\
&&241.7&&1.872\,(2)&1.857\,(2)&1.887\,(2)&  -8.003\,(12)\\
&&254.6&&1.875\,(2)&1.865\,(2)&1.885\,(2)&  -5.121\,( 7)\\
&&256.4&&1.873\,(2)&1.865\,(2)&1.882\,(2)&  -4.411\,( 7)\\
       30\,Jun\,2014&        07:30--08:13&321.0&1.0&2.054\,(1)&2.119\,(2)&1.988\,(2)&  32.973\,(37)\\
&&322.8&&2.057\,(1)&2.120\,(2)&1.991\,(2)&  32.389\,(36)\\
&&331.6&&2.061\,(1)&2.127\,(2)&1.993\,(1)&  33.664\,(33)\\
&&333.5&&2.062\,(1)&2.128\,(1)&1.994\,(1)&  33.808\,(32)\\
       18\,Jul\,2014&        03:49--04:14&279.2&1.2&1.806\,(2)&1.805\,(2)&1.806\,(2)&  -0.526\,( 1)\\
&&281.1&&1.807\,(2)&1.803\,(2)&1.811\,(2)&  -2.071\,( 3)\\
&&290.0&&1.818\,(2)&1.817\,(2)&1.819\,(2)&  -0.632\,( 1)\\
&&291.8&&1.820\,(2)&1.820\,(2)&1.820\,(2)&   0.000\,( 0)\\
       21\,Jul\,2014&        04:23--04:43& 88.3&2.5&1.614\,(1)&1.593\,(1)&1.635\,(1)& -12.722\,(11)\\
&& 90.2&&1.613\,(1)&1.592\,(1)&1.634\,(1)& -13.002\,(10)\\
&& 99.0&&1.617\,(1)&1.591\,(1)&1.644\,(1)& -15.909\,(15)\\
&&100.7&&1.617\,(1)&1.592\,(1)&1.642\,(1)& -15.104\,(13)\\
       29\,Jul\,2014&        04:55--05:12&240.2&1.4&1.733\,(1)&1.739\,(1)&1.726\,(2)&   3.591\,( 4)\\
&&241.7&&1.734\,(2)&1.742\,(2)&1.725\,(2)&   4.869\,( 7)\\
&&254.6&&1.756\,(2)&1.760\,(2)&1.751\,(2)&   2.598\,( 3)\\
&&256.4&&1.757\,(2)&1.762\,(2)&1.752\,(2)&   2.682\,( 4)\\
       26\,Aug\,2014&        23:28--23:48& 88.3&3.4&1.523\,(1)&1.542\,(1)&1.508\,(1)&  11.207\,( 8)\\
&& 90.2&&1.520\,(1)&1.540\,(1)&1.503\,(1)&  12.276\,( 7)\\
&& 99.0&&1.512\,(1)&1.547\,(1)&1.477\,(1)&  23.589\,(17)\\
&&100.7&&1.512\,(1)&1.550\,(1)&1.475\,(1)&  25.422\,(16)\\
\hline 
\end{longtable} 
\end{center}


\section{Monte Carlo tests}
\label{MonteCarlo}

We have performed a Monte Carlo analysis of the robustness of our $RM$ estimates. In this section, we show how the null hypothesis of no $RM$ detection can be rejected, given our observations. 

For this analysis, we have generated random values of $R_{pol}$ at the same parallactic angles of our observations, and have fitted these random data using Eq. \ref{FarRotEq3}. The random values have been generated using either a Gaussian distribution (with zero average and $\sigma$ given by half of the maximum $R_{pol}$ observed) or a uniform distribution (with extreme values equal, in absolute value, to the maximum $R_{pol}$ observed). The choice of either random distribution does not affect the conclusions here reported. 

\begin{figure}[ht!]
\centering
\includegraphics[width=10cm]{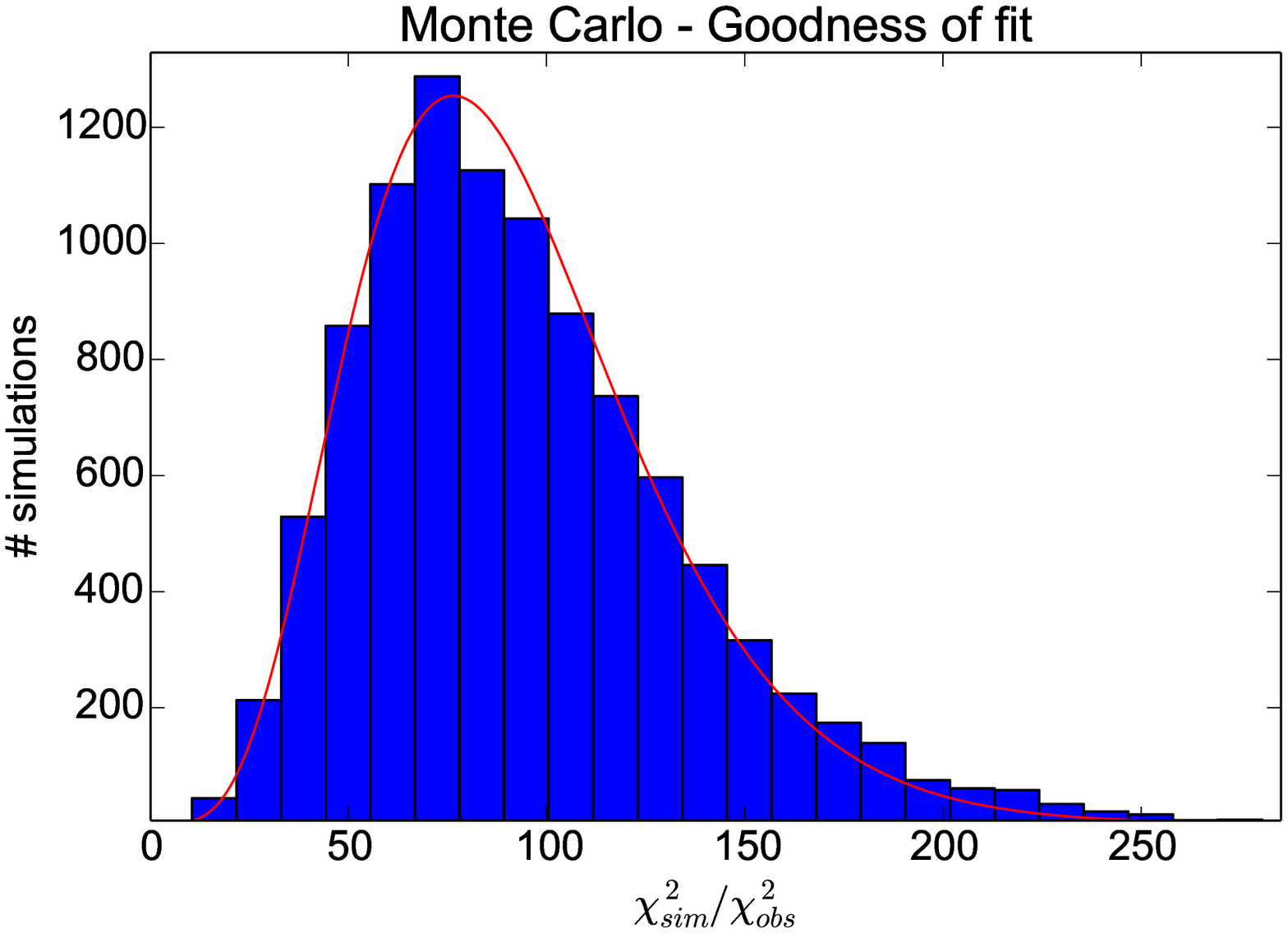}
\caption{Histogram of the best-fit $\chi^2$ values obtained from our Monte Carlo analysis of the null hypothesis of $RM$ detection. The parallactic angle coverage is the same as that of our observations on 5 May 2014. The $\chi^2$ values are scaled to the value obtained from the observations. The red line is a scaled version of the expected $\chi^2$ probability density function.}
\label{MonteCarloFig}
\end{figure}

In Fig. \ref{MonteCarloFig}, we show, as an example of our simulation results, the histogram of minimum $\chi^2$ values resulting from the fits to $10^4$ random datasets, with the same parallactic-angle coverage as that of our observations on 5 May 2014. The $\chi^2$ values are scaled to that from our observed data. The red line corresponds to the expected $\chi^2$ distribution with 12 degrees of freedom (i.e., 15 data points minus 3 fitting parameters), which we have scaled in the X axis (to account for our normalization of the $\chi^2$) and in the Y axis (to account for the total number of simulations). The Monte Carlo histogram roughly follows a random $\chi^2$ distribution, as expected. We notice that the goodness of fit from our observations {\em is much higher than the highest one obtained from the random data}. Indeed, according to the $\chi^2$ distribution shown in Fig. \ref{MonteCarloFig} (red line), the probability that a random dataset allows us, just by pure chance, to get a fit as good as that from our observations is lower than $10^{-10}$.

\begin{figure}[ht!]
\centering
\includegraphics[width=10cm]{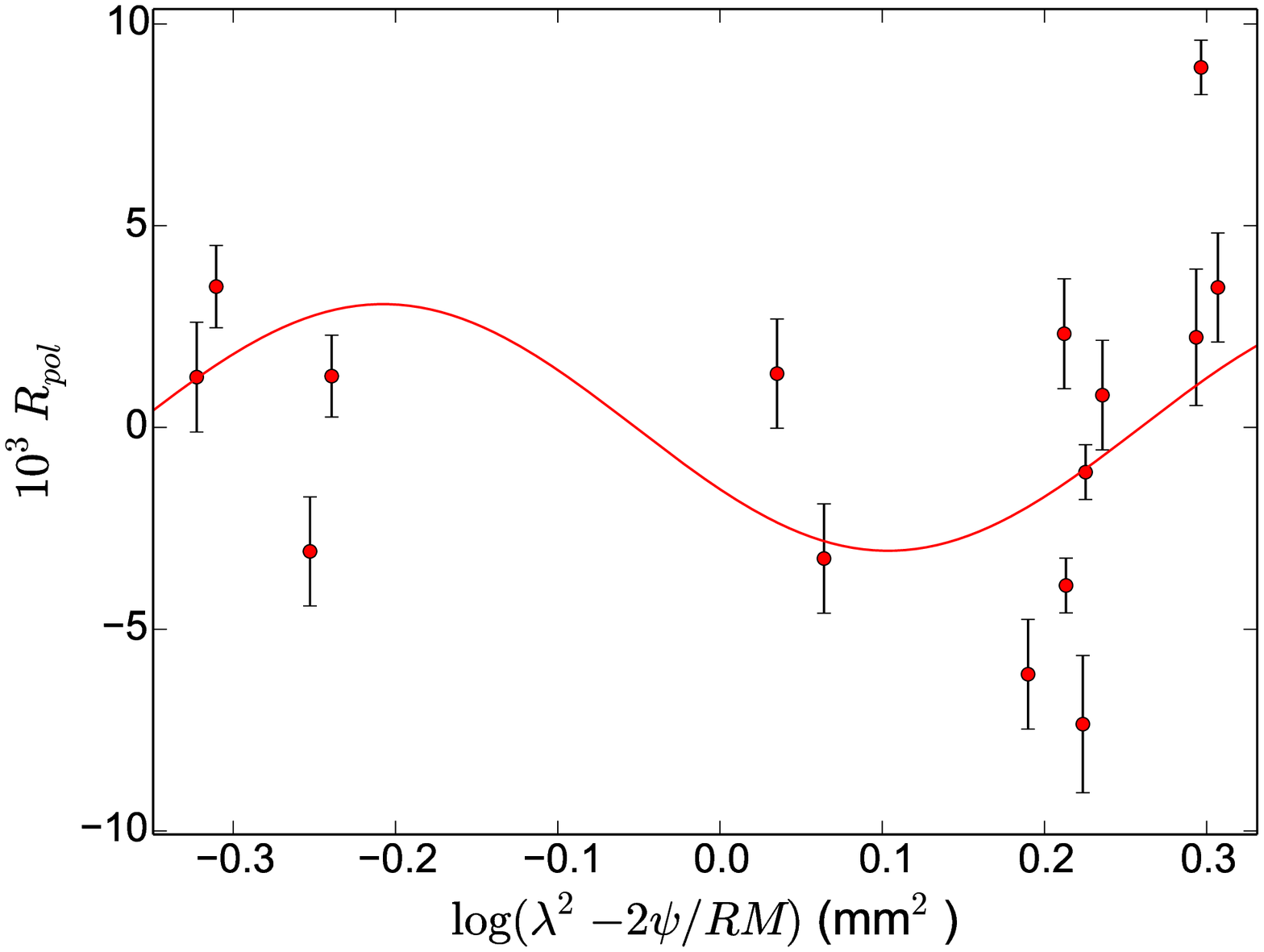}
\caption{Our best fit to random values of $R_{pol}$, selected from a set of $10^4$ simulations. The parallactic angle coverage is the same as that of our observations on 5 May 2014.}
\label{BestSimul}
\end{figure}

For completeness, we show in Fig. \ref{BestSimul} the best fit obtained from our full set of $10^4$ random trials. It can be readily seen that the quality of this fit is far worse than the results obtained from our observations.

\end{document}